\documentclass[12pt]{article}
\usepackage[english]{babel}
\usepackage[utf8x]{inputenc}
\usepackage{amsmath}
\usepackage{graphicx}
\usepackage{etoolbox}
\usepackage{changepage}
\usepackage{titlesec}
\usepackage[parfill]{parskip}
\usepackage[margin=1in]{geometry}
\usepackage{times}
\usepackage{titlesec}
\usepackage{float}
\usepackage[numbers,sort&compress]{natbib}

\usepackage{booktabs}
\usepackage{array}
\usepackage{subcaption}

\titleformat*{\section}{\normalsize\bfseries} 
\titleformat*{\subsection}{\normalsize\itshape} 

\title{\large \textbf{Educating a Responsible AI Workforce: Piloting a Curricular Module on AI Policy in a Graduate Machine Learning Course}} 

\author{\normalsize James Weichert\\
\normalsize jamesweichert@vt.edu\\
\normalsize Department of Computer Science\\\
\normalsize Virginia Tech
\and
\normalsize Hoda Eldardiry\\
\normalsize hdardiry@vt.edu\\
\normalsize Department of Computer Science\\
\normalsize Virginia Tech
}

\date{} 

\makeatletter 
\patchcmd{\@maketitle}{\begin{center}}{\begin{adjustwidth}{0.5in}{0.5in}\begin{center}}{}{}
\patchcmd{\@maketitle}{\end{center}}{\end{center}\end{adjustwidth}}{}{}
\makeatother

\begin{document}
\raggedright
\maketitle
\thispagestyle{empty}
\pagestyle{empty}

\section*{Abstract}

As artificial intelligence (AI) technologies begin to permeate diverse fields—from healthcare to education—consumers, researchers and policymakers are increasingly raising concerns about whether and how AI is regulated. It is therefore reasonable to anticipate that alignment with principles of `ethical' or `responsible' AI, as well as compliance with law and policy, will form an increasingly important part of AI development. Yet, for the most part, the conventional computer science curriculum is ill-equipped to prepare students for these challenges. To this end, we seek to explore how new educational content related to AI ethics and AI policy can be integrated into both ethics- and technical-focused courses. This paper describes a two-lecture \textit{AI policy module} that was piloted in a graduate-level introductory machine learning course in 2024. The module, which includes an in-class active learning game, is evaluated using data from student surveys before and after the lectures, and pedagogical motivations and considerations are discussed. We find that the module is successful in engaging otherwise technically-oriented students on the topic of AI policy, increasing student awareness of the social impacts of a variety of AI technologies and developing student interest in the field of AI regulation.


\section*{Introduction}

The explosive growth of artificial intelligence (AI) technologies is widely documented and increasingly evident in everyday life: some responses from the search engine Google now include an ``AI Overview'' inserted before the first webpage link; companies like Tesla and Waymo have seen success in implementing partial or full autonomous driving in vehicles on live roads; and ``Apple Intelligence'' was the flagship feature for the launch of Apple's new smartphone in fall 2024. Yet what legal or policy response this technological growth will precipitate is less certain \cite{jobin_global_2019, parinandi_investigating_2024}. Nevertheless, it should be expected that the development and enactment of regulatory frameworks for AI will demand AI engineers with a command not only of the technical intricacies of AI models, but also of the policy and regulatory landscape for AI development and compliance \cite{aler_tubella_how_2023}. This is made clear by the 2023 U.S. Executive Order on \textit{Safe, Secure, and Trustworthy Development and Use of Artificial Intelligence} \cite{exec_order}, which calls for an “AI talent surge” as well as the training of the federal workforce on “AI issues…as well as relevant policy, managerial, procurement, regulatory, ethical, governance, and legal fields,”

Our previous research has found that although some students studying AI are interested in career paths related to AI policy, only a third of students surveyed thought their computer science (CS) courses adequately prepared them for these career options. This conclusion is supported by a review \cite{weichert_i_2025} of computing ethics requirements for 250 CS undergraduate programs worldwide, which found that only one third of programs required students to take a computing ethics course in order to graduate, while nearly one half of programs did not offer any computing-related ethics courses. When students do take an ethics course, they are usually presented with outdated course material focusing on more `traditional' CS ethics topics such as privacy or intellectual property \cite{weichert_eval}. Even if content on AI is present, ethical implications tend to be discussed only in the abstract, omitting discussion on the importance of regulation and governance of AI.

We see curricular content on AI ethics and policy as valuable in two respects. Not only may CS students be themselves interested in policy-related careers, but the lack of instruction on how to apply (operationalize) ethical principles and regulatory requirements to the design and deployment of AI technologies represents, in our opinion, a ‘missing middle’ between technical computer science and policy-focused (e.g. public policy) curricula. Thus, the integration into CS programs of practical content related to ensuring regulatory compliance with AI systems is useful to the latter segment of students, while more open-ended discussions about AI ethics and policy may, on the other hand, inspire CS students to explore policy-related career paths. 

In order to investigate how this content integration can be achieved, we set out to develop a two-lecture `curricular module' on AI policy. This module was then piloted at the end of a graduate-level introductory machine learning course (i.e. ``ML 1'') at a large public university in the U.S. during fall 2024. These two lectures replaced and expanded upon the one lecture in the course previously reserved for discussion of ethics and AI. In order to better engage students in an otherwise `technical' course, the module incorporates a mix of lecture-style presentation, class discussion, and an in-class role-playing game inspired by the popular party game \textit{Mafia}. This active learning \cite{bonwell_active_1991, freeman} component is key for diversifying away from the lecture-heavy approach taken in many computing ethics courses. The content of the two classes is described further in the \textit{AI Policy Module} section. In order to gauge the effectiveness of the module, students completed pre- and post-module surveys \cite{vogt, shaw_knowing_2019} which aimed to assess student interest in AI policy and measure competencies related to the application of policy to technical AI development. 

The overall goals of this research are twofold. First, this pilot may provide other instructors with inspiration for content or activities to include in similar ‘modules’ in their own courses. The purpose of this pilot module, then, is to demonstrate that discussions around AI ethics and policy can easily and effectively be incorporated into existing technical courses. As a secondary objective left for future work, the pre- and post-surveys will help to further evaluate student interest in AI policy. A larger survey sample may also help to identify competencies that may be relevant to AI practitioners tasked with ensuring compliance with AI regulation and ‘responsible AI’ (RAI) principles, which are often abstract and sometimes contradictory. The latter is especially useful to policymakers and educational administrators who must adapt university curricula to prepare students for the AI workforce, which will increasingly involve ethical, social, and regulatory dimensions in addition to technical ones. 

\section*{Literature Review}

\subsection*{Computing Ethics Education}

Surveys of ethics education in the computing or computer science context have identified significant diversity in \textit{what} content is taught, \textit{how} it is taught, and \textit{to whom} it is taught. Our systematic review \cite{weichert_i_2025} of undergraduate CS degree requirements at 250 universities worldwide revealed that standalone computing ethics courses are a required part of the degree for only one third of programs, while nearly half of programs do not offer a computing ethics course at all. Thus, the reach of CS ethics education is perhaps less extensive than commonly thought. However, as Brown et al. \cite{brown_teaching_2024} show in their systematic review of 100 CS ethics education research papers published in top venues, there is a roughly even distribution of ethics teaching between standalone courses (32\%) and integration of ethics content into one (26\%) or more (35\%) modules of a technical course. With respect to teaching approach, instructors use a mix of \textit{pedagogical strategies} in delivering content, the most common of which are class discussions, readings, lectures, and writing assignments \cite{brown_teaching_2024}. Active learning techniques such as role-playing simulations, debates, and games are less common.

While approaches to teaching computing ethics vary across institutions, what is taught under the umbrella of `computing ethics' varies significantly more. A review of 115 `tech ethics' course syllabi by Fiesler et al. \cite{fiesler_what_2020} identified 15 course topics each featured in over 10 courses, ranging from \textit{professional ethics} to \textit{inequality, justice \& human rights} to \textit{cybersecurity}. Among the 15 topics, \textit{law \& policy} was most prevalent, appearing in 57\% of syllabi, while \textit{AI \& algorithms} appeared in 48\%. However, our review \cite{weichert_eval} of 116 computer science ethics courses (having only 16\% overlap with the courses in the Fiesler et al. corpus) concludes that a shift from teaching about \textit{AI as an ethics case study} to \textit{AI ethics} as a subfield—where emphasis is placed not only on the impacts of AI technology, but equally on how to facilitate the responsible use and ethical governance of AI—is only beginning to take place. The nascency of AI ethics and policy education provides ample opportunity for experimentation in content (e.g. topics, case studies, etc.), format (lecture vs. seminar vs. lab), and pedagogical strategies (e.g. discussions, writing assignments, in-class activities, etc.).

\subsection*{AI Ethics and Policy}

As AI systems feature in more consumer technologies, and as student interest in AI and machine learning continues to grow, the CS education community has coalesced around a call to develop ethics education tailored to the particular impacts and challenges posed by AI \cite{burton_ethical_2017, furey_ai_2019, garrett_more_2020, borenstein_emerging_2021, raji_you_2021}. What shape this AI ethics curriculum will take, however, is still largely unclear. The absence of a solidified `canon' for AI ethics teaching is conspicuous in the variety of case studies and perspectives used when discussing AI in ethics courses \cite{weichert_eval}. A few core topics and similar case studies do emerge, however. Through their analysis of 51 ethics and technical AI courses, Garrett et al. \cite{garrett_more_2020} identify \textit{bias}, \textit{fairness}, \textit{privacy}, and \textit{automation} as the most common themes through which the social impacts of AI are evaluated. For example, the authors note many courses use the COMPAS automated recidivism algorithm \cite{larson_how_2016} as a case study for discussing AI bias. We posit that these themes are most common because they represent adaptations of themes common in the computing ethics curriculum before the advent of AI. Privacy and the impact of automation on employment have been mainstays on syllabi for decades, and although the impacts of algorithmic bias have certainly become more evident in recent years, COMPAS (released in 2012) is better conceptualized as a \textit{statistical} algorithm than an \textit{AI} algorithm in the contemporary sense. Impacts more unique to AI systems, such as hallucination or the impact of massive computing demands on the environment, have yet to see widespread adoption in the AI ethics curriculum.

As we note above, \textit{AI policy} is even less common as a topic in existing ethics courses, and syllabi revisions are only recently beginning to show a shift from thinking about \textit{potential} impacts of AI to \textit{actual} real-world impacts, as well as how AI should be overseen to mitigate harms in the future \cite{weichert_eval}. On the one hand, the lack of inclusion in ethics courses is understandable given that the landscape of AI regulation is still in its infancy. Whereas data privacy, for example, offers students concrete regulatory case studies like the U.S. Patriot Act or (more recently) the European Union's General Data Protection Regulation (GDPR), it will take a few years to evaluate the effect of AI legislation just now coming into effect. Nevertheless, the horizon of AI regulation among key state actors (United States, China, European Union) is more or less clear~\cite{schiff_whats_2020, weichert_policies}. The integration of a forward-looking perspective on AI policy as part of an AI ethics curriculum would therefore provide students with valuable context as they embark on careers that will increasingly expect AI engineers to navigate policy requirements. Given that, to our knowledge, this approach has not been previously investigated, we made the intentional decision to develop our curricular module around \textit{AI ethics} and \textit{AI policy} as dual foci. 

\section*{Methodology}

\subsection*{Study Design}

The \textit{AI policy module} described below is designed to integrate discussions on ethics and social implications of AI into a `technical' AI/ML course, providing students with opportunities to connect the methods and algorithms learned in class with `real-life' impacts. As such, we piloted the module at the end of a graduate-level machine learning course (``ML 1''). ML 1 covers a variety of foundational topics in machine learning, including supervised learning (regression, classification), unsupervised learning (clustering, anomaly detection), and neural networks.

The two lectures in the module were presented at the end of the semester, after the course's final exam but the week before students gave presentations on their course-long group projects. This scheduling had two advantages: (1) the module could draw on the entirety of the course's technical material, and (2) students could focus on the module instead of worrying about studying for their final exam. The module was led by the first author of this paper, who was the course's graduate teaching assistant and whose research focuses on AI ethics and policy education.

\subsection*{Participants}

Although three-fourths of the students enrolled in ML 1 were graduate students in computer science, the course was open to all graduate students interested in machine learning and also included students from other engineering and science disciplines, including civil engineering, industrial engineering, and psychology. This disciplinary diversity, combined with the diversity in research interests among students, contributed to a variety of perspectives during the module's class discussions. Each of the two lectures in the module was attended by between 20 and 30 students (out of a total enrollment of 40). This represented a significant drop in attendance for the course, attributable to the fact that module material was not being assessed and class attendance was not tracked. While the smaller class size paralleled a graduate seminar environment, the lower attendance is nevertheless a concern we address in the \textit{Discussion} section.

\subsection*{Data Collection}

The primary source of data, both quantitative and qualitative, for evaluating the module came from student responses to a pre-module and a post-module survey. The \textit{Pre-Survey} was administered at the beginning of the first lecture after introducing the aim of the module, while the \textit{Post-Survey} was administered at the end of the second lecture. In order to encourage honest feedback from students, no identifying information was collected, leading to lower response rates ($n = 17$ for the \textit{Pre-Survey} and $n = 13$ for the \textit{Post-Survey}) despite time being allocated during the lectures for students to complete the surveys. Both surveys include a section each on ``AI Ethics'' and ``AI Policy'', corresponding to the module's two lecture topics. The questions in these sections consist of short free-response and 1-5 Likert scale questions, which are identical between the two surveys, enabling a comparison of student attitudes before and after the module. The \textit{Post-Survey} contains an additional section for students to evaluate the AI policy module as a whole. Appendix A summarizes each survey question.

Given the relatively small class size and pilot nature of the project, small sample sizes were expected. As such, in evaluating the module and drawing conclusions, we use broader trends in the survey data and student testimonials to support our qualitative observations. We present these results in the \textit{Evaluation} section. Pedagogical considerations and reflections from the point of view of the instructor are included in the \textit{Discussion} section. 

\section*{AI Policy Module}

The two lecture slots planned for the module help to split \textit{AI ethics} and \textit{AI policy} into their own distinct lectures. In the first class period, the focus on AI ethics transitions students from thinking about the details of machine learning algorithms to thinking about the impact of AI on society. The second lecture on AI policy then shifts focus from \textit{thinking} about AI's impacts to \textit{acting} responsibly with AI. The structure and core content of each lecture are outlined below.

\subsection*{Lecture 1: ``Ethics and Artificial Intelligence''}

\begin{figure}
    \centering
    \begin{subfigure}{.47\textwidth}
        \includegraphics[width=1\linewidth]{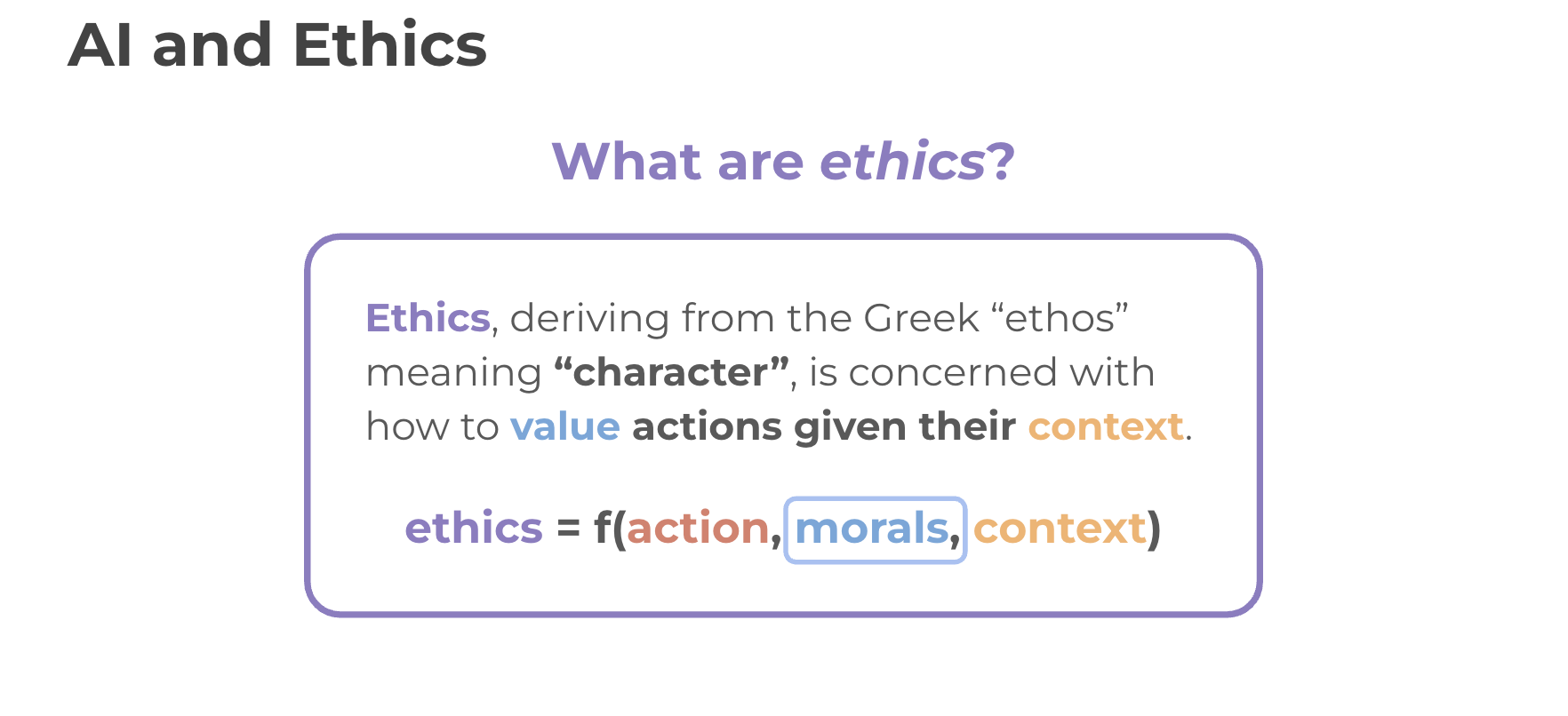}
        \caption{Slide from the ``Ethics and Artificial Intelligence'' lecture formulating \textit{ethics} as a function of an action, its context, and morals.}
        \label{fig:ethics-def}
    \end{subfigure}%
    \hspace{2em}%
    \begin{subfigure}{.47\textwidth}
        \includegraphics[width=1\linewidth]{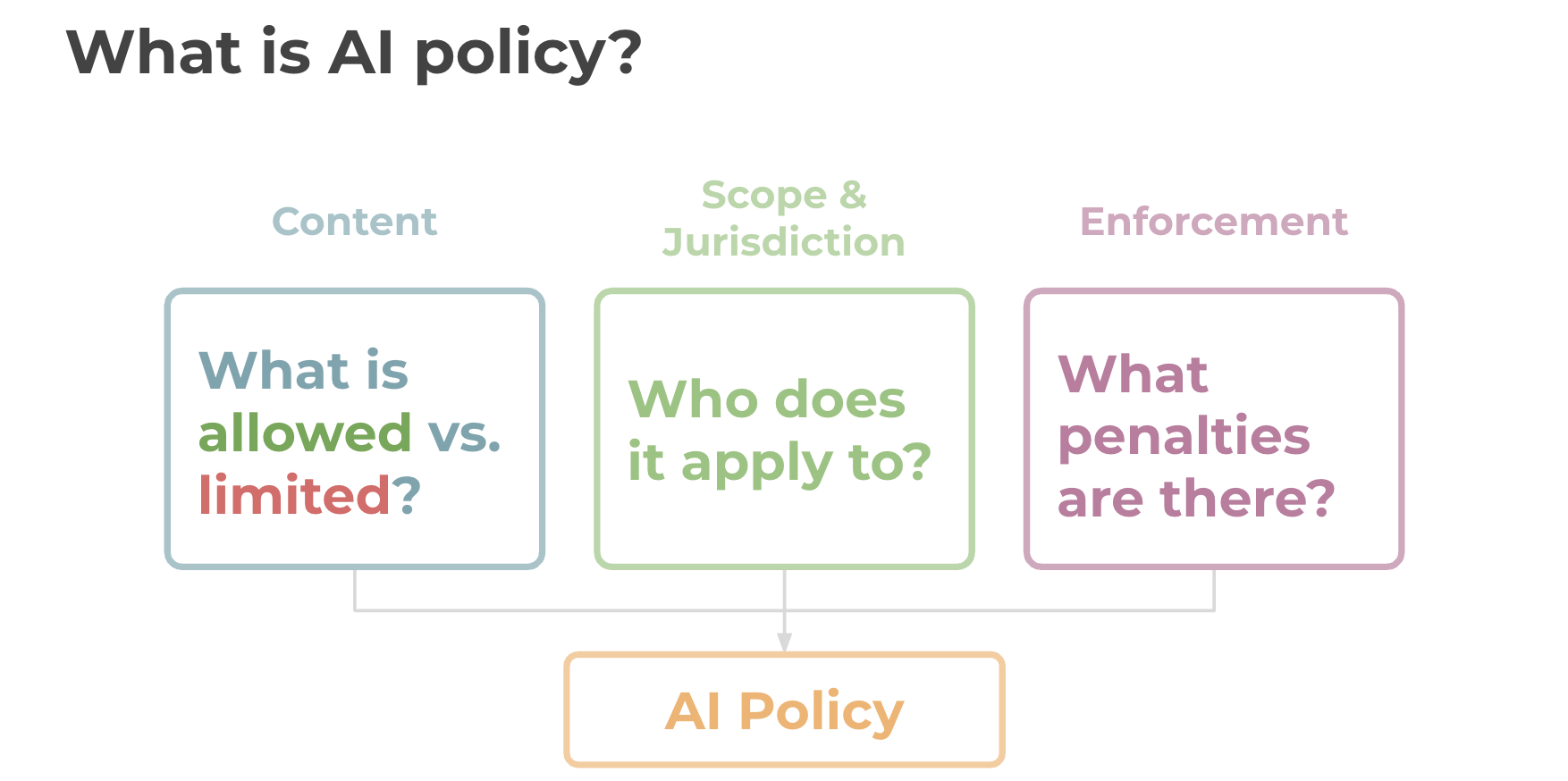}
        \caption{Slide from the ``Regulating Artificial Intelligence'' lecture deconstructing \textit{AI policy} into content, scope \& jurisdiction, and enforcement.}
        \label{fig:ai-policy}
    \end{subfigure}
    \caption{}
\end{figure}

The first class period in the module contains the most conventional lecture-style presenting of the two lectures, necessary for establishing a common understanding of how to conceptualize the social impacts of AI. This includes establishing a common definition of ``ethics'', how individual morality differs from collectively-held ethics, and how specific ethical frameworks are declaratively defined through ethical principles. However, throughout this introduction to ethical theory, we attempt to provide more concrete formulations for abstract concepts and to place the theory in the context of AI. Figure \ref{fig:ethics-def}, for instance, shows a quasi-mathematical formulation for the term \textit{ethics} as a function of an action, its context, and a particular moral framework.

In addition to class discussions prompted by open-ended questions (e.g., ``How should we reconcile competing ethical principles?''), a key element of student engagement in this lecture is achieved through three case studies to highlight specific ethical principles and their relevance to AI. The first principle of \textit{nonmaleficence} is framed in the context of autonomous vehicles using the example ``How (not) to get hit by a self-driving car,'' an interactive street-based game by Tomo Kihara~\cite{noauthor_playable_nodate} where players try to move such that an AI pedestrian detection system does \textit{not} detect them crossing the street. Then, students are asked to put themselves in the shoes of self-driving car developers in deciding how to prioritize some lives over others by choosing between two options for a given car-pedestrian collision scenario from the MIT Moral Machine~\cite{awad_moral_2018}. The case study is left open ended with the question: ``how do you put nonmaleficence into practice with autonomous vehicles?'' when there is no way to avoid \textit{some} harm.

The second principle of \textit{privacy} is connected with the algorithmic recommendation system of the social media platform TikTok and the controversy surrounding its prohibition in the United States. After establishing students' attitudes towards TikTok's collection of user data, this case study encourages a broader discussion about whether consumers in the age of ``Big Data'' are really concerned about \textit{privacy} or rather about what can be \textit{inferred} from collected data. Finally, the ability of some AI systems (e.g. large language models, video generators, etc.) to generate convincing synthetic content (``deepfakes") or provide false information (``hallucination") is used to explore the principle of \textit{trustworthiness}. First, a political campaign advertisement featuring a video deepfake of former Pakistani prime minister Imran Khan—produced while Khan was in jail—provides a counterexample to the notion that synthetic AI content is \textit{necessarily} bad \cite{haq_deepfakes}. Then, students are encouraged to try to `hack' OpenAI's large language model ChatGPT so that it hallucinates, malfunctions, or gives an inappropriate response.

The first lecture concludes by asking students to subjectively rank the three highlighted principles—plus three others (\textit{transparency}, \textit{justice} and \textit{autonomy}) which together summarize Floridi and Cowl's ``Unified Framework for AI in Society'' \cite{floridi_unified_2019}—in terms of their relative importance. Students' rankings are contrasted with the appearance frequency of these principles in the 84 public and private sector AI ethics guidelines reviewed by Jobin et al. \cite{jobin_global_2019}. This juxtaposition highlights how different ethical frameworks (e.g. Floridi and Cowl's ``Unified Framework,'' the White House's ``Blueprint for an AI Bill of Rights,'' Microsoft's ``Responsible AI Standard'') can emphasize different principles and therefore lead to different design decisions. Finally, the question of which principles take precedence and how principles are translated to practices \cite{kim_exploring_2023} is framed as the motivation for \textit{AI policy}, to be covered in the second lecture.

\subsection*{Lecture 2: ``Regulating Artificial Intelligence''}

The second lecture in the module frames \textit{AI policy} as the medium through which principles in AI ethics frameworks are put into practice with code and institutional structures. The goal of this lecture is to introduce students to the components of policy, influences on how governments and corporations shape policy, and key nascent government AI regulation. 

First, ``policy'' is deconstructed into three components, as shown in Figure \ref{fig:ai-policy}: \textit{content} (what is allowed vs. limited?), \textit{scope \& jurisdiction} (who does it apply to?), and \textit{enforcement} (what penalties are there?). The motivations for regulating AI are discussed from the perspectives of both the private and public sectors. AI companies may be influenced by consumer opinion calling for AI regulation \cite{auld_governing_2022}, or may engage in what Floridi \cite{floridi_translating_2019} calls \textit{ethics shopping} and \textit{bluewashing} to retrofit ``ethical frameworks to justify their current behaviors'' or use self-regulation ``in order to lobby against the introduction of legal norms,'' respectively. Like private companies, governments might act to regulate AI in order to protect consumer safety and public opinion. President Biden's executive order on AI \cite{exec_order}, for example, notes that ``mistakes by or misuse of AI could harm patients, cost consumers or small businesses, or jeopardize safety or rights," However, lawmakers may likewise act for less altruistic reasons, including preserving competitive economic advantage over other nations \cite{schiff_whats_2020}. Moreover, AI regulation is prone to be influences by ideological differences \cite{parinandi_investigating_2024} and lobbying by AI companies themselves \cite{floridi_translating_2019}.

To round out the module's introduction on AI policy, we provide a short overview of enacted or pending government AI regulation among the three primary governmental actors when it comes to AI: China, the European Union, and the United States~\cite{weichert_policies}. Taken together, this background helps to establish important context that students will use in simulating the process of formulating, lobbying for or against, and passing AI legislation in the primary activity for the second lecture, the in-class game \textit{Congress vs. Evil Inc.}, which we describe below.

\subsection*{In-Class Game: Congress vs. Evil Inc.}

The limited class time and desire not to ask students to prepare outside of class (e.g., for a debate) presented considerable constraints on what kind of active learning activity could be designed to connect with the AI policy content of the second lecture. The result of wanting to recreate the conflicting interests involved in the policymaking process was a game to simulate the U.S. Congress' role in AI regulation, which we call \textit{Congress vs. Evil Inc.} In this dramatized and hyperbolized simulation, inspired by the popular party game \textit{Mafia} (also called \textit{Werewolf}), players (students) assume one of three roles: (1) leaders of \textit{Evil Inc.}, a fictional multinational technology company; (2) members of Congress charged with oversight and regulation of AI; and (3) voters, who want to induce Congress to pass robust (but not overly restrictive) AI regulation through their voting patterns. The goals and player mechanics for each role are summarized in Table \ref{tab:roles}.

\begin{table}[]
    \centering
    \begin{tabular}{l|p{4cm}|p{4cm}|p{4cm}}
         & \multicolumn{1}{c|}{\textbf{Voters}} & \multicolumn{1}{c|}{\textbf{US Congress}} & \multicolumn{1}{c}{\textbf{Evil Inc.}} \\
         \midrule
         \textbf{Goal} & Robust AI regulation, good user experience & Political survival & Profit \\
         \midrule
         \textbf{Win Condition} & Key regulation is adopted & Stay in Congress until the end of the game & Prevent the adoption of key regulation \\
         \midrule
         \textbf{Action} & Can, if desired, vote out one member of Congress per turn & During the Congressional hearing, make one statement or pose one question to the \textit{Evil Inc.} CEO & Can, if desired, bribe members of Congress \\ 
    \end{tabular}
    \caption{Goal, win condition and available action for each player in \textit{Congress vs. Evil Inc.}}
    \label{tab:roles}
\end{table}

\newpage

\textit{Congress vs. Evil Inc.} takes the form of a turn-based game narrated by the instructor following a predefined scenario inspired by real-world regulatory efforts. For example, Scenario 3 on `general-purpose' AI is inspired by the EU AI Act's requirement \cite{eu_ai_act} that proprietors of `general purpose' AI systems report details about model architecture and training processes to national AI authorities. Each scenario, along with the proposed AI regulations that Congress can vote on, are described in Appendix B. To start the game, only members of \textit{Evil Inc.} are told internal company information that motivates their lobbying efforts. For example: ``\textit{Evil Inc.}'s large language model is only successful because its model architecture is kept a secret, so \textit{Evil Inc.} should prevent Congress from requiring the disclosure of any model architecture information,'' Each round, \textit{Evil Inc.} members decide how to distribute a limited number of `bribe cards' among members of Congress, representing the company's lobbying capital. Each card has a dollar amount which, if larger than the `bribe threshold' on a member of Congress' player card, requires that member to vote \textit{against} the pending AI regulation. Knowing whether or not they were successfully bribed, each member of Congress then has the opportunity to ask a question of the \textit{Evil Inc.} CEO (the instructor) or make a statement during the congressional hearing. Then, the members of Congress vote on the pending regulation, which is enacted if it receives a majority of votes and fails otherwise. Finally, based on the public statements of the members of Congress and their voting records, the voters can decide to vote out up to one member of Congress per turn if they are unhappy with their representative's legislative actions. This cycle repeats for each of the proposed AI regulations provided for the scenario. If, in the end, the key regulation is not adopted then \textit{Evil Inc.} is said to have won, whereas if the regulation is adopted, then the voters have won. Mimicking their desire for reelection, members of Congress win regardless of the regulatory outcome if they survive in Congress without being voted out. 

In a very limited playing time and with no prior preparation by the players, \textit{Evil Inc.} aims to emulate the competing forces at play in the real-world AI policy landscape. Each role has an ability to `check' another, leading to excitement, intrigue, and (hopefully) balanced gameplay. Important game design choices and reflections are included in the \textit{Discussion} section.  


\section*{Evaluation}

Below, we summarize our findings with respect to the impact of the AI policy module on students' attitudes and competencies related to AI ethics policy. As previously noted, the small sample sizes of our pre- and post-module surveys ($n=17$ and $n=13$, respectively) necessarily limit the robustness of the conclusions that can be drawn from our quantitative data. However, given the pilot nature of this project, we regard these survey data as nevertheless helpful to broadly assess the efficacy and impact of the module in this early stage of development.

\subsection*{Student Attitudes and Competencies}

Unsurprisingly, student \textit{attitudes} towards issues of AI ethics and policy remain largely unchanged after the two lectures. Figures \ref{fig:ethics-likert} and \ref{fig:policy-likert} in Appendix C show the changes in the responses between the two surveys for ethics- and policy-related questions, respectively. Overall, the Likert-scale distributions change little, especially when only considering overall agreement versus disagreement. We view this as evidence not of any deficiency in our module, but rather as support for the hypothesis that CS students develop their attitudes toward AI mainly through their own exploration and use of AI tools, both inside and outside of the classroom. The process of ethical development among students in the context of AI therefore merits further investigation.

\begin{figure}
    \centering
    \includegraphics[width=0.5\linewidth]{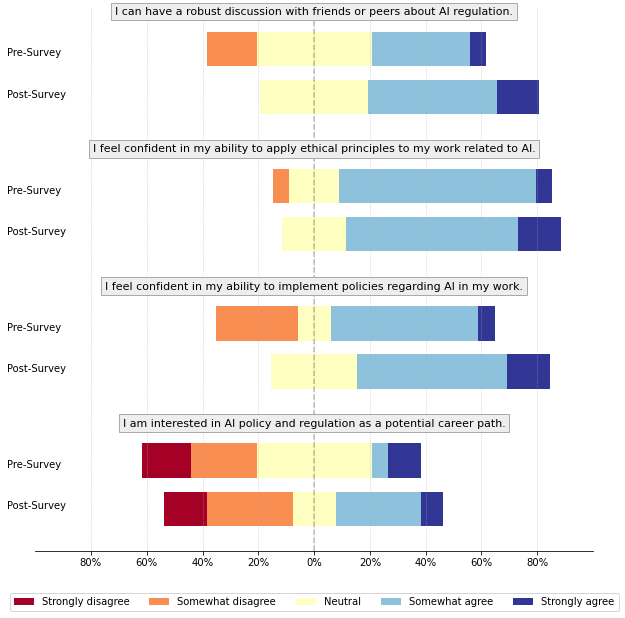}
    \caption{Student self-assessment of AI policy competencies before and after the \textit{AI policy module}.}
    \label{fig:comp-likert}
\end{figure}

However, there was a noticeable increase in the Post-Survey of students who indicated that they intend to follow news about government regulation of technology and AI going forward. We view this as an indication that the \textit{policy} component of the module, presented the students with a new and useful perspective on the AI technologies with which they were already familiar. This hypothesis is supported by students' responses to the AI policy competency questions, shown in Figure \ref{fig:comp-likert}. Following the two lectures, students felt more confident in their ability to have a discussion with peers about AI regulation (62\% strongly agree or somewhat agree, compared to 41\% in the pre-survey) and implement AI policies in their work (69\% vs. 59\%). Finally, the module seems to cultivate an interest in the field of AI policy among a few students, with the percentage of students who expressed interest in AI policy and regulation as a potential career path increasing from 18\% to 38\%. The Post-Survey percentage is close to the 44\% of students expressing interest in AI policy careers in our previous study \cite{weichert_computer_2024} and suggests that interest in the field can be broadened through exposure to AI policy content in computing courses.

\subsection*{Student Evaluations of the AI Policy Module}

Student feedback on the module overall consistently positive, with 92\% of respondents strongly agreeing or somewhat agreeing that ``the content of the lectures was interesting,'' ``the lectures were engaging and interactive,'' and ``the lectures were a good use of class time''. Moreover, a majority of respondents strongly agreed that they liked ``the level of interactivity in these lectures compared to previous lectures,'' Optional written feedback highlighted that the lectures were ``engaging,'' ``interesting,'' ``fun,'' and ``informative''. One student noted that ``I would take a whole class on [this topic] if offered''. Among the two students who listed a favorite part of the module, the in-class game was mentioned both times. However, comments suggesting that the incentives for different roles in \textit{Congress vs. Evil Inc.} should be adjusted indicate that the game's mechanics need to be refined to ensure that all roles are balanced and equally engaging to play.


\section*{Discussion}

\subsection*{Module Content}

The primary strength of this one-week module is that it is short enough to be added to an existing technical course without having to remove much existing content, yet still long enough to serve as a meaningful introduction to the social impacts of AI. The dual focal points of \textit{AI ethics} and \textit{AI policy} divide cleanly between the two lectures, while the latter smoothly transitions from (and extends on) the former. In the first lecture, the focus is on how to conceptualize—\textit{think} about—the impacts of AI, while in the second, the focus evolves into how to \textit{act} to achieve `ethical AI' through regulation. The second lecture completes a progression from ethical principle to practice \cite{kim_exploring_2023}, which is first achieved through the formulation of a framework of ethical principles, then through the development of policy and enactment of regulation through code, governance structures, and institutional culture.

If the purpose of this module is distilled down to one goal, it is to introduce students to the landscape of AI policy in such a way as to enable them to navigate the interplay of AI regulation on their work if and when it arises. In other words, we aim to impart the context and `vocabulary' necessary to discuss or take part in AI policymaking decisions. As our survey data show, it is not reasonable to expect all students to find genuine academic or professional interest in technology policy. For this set of students, the end of the module is the end of their engagement with the topic, having at least become familiar with the basics of AI ethics and policy. But for some subset of students, adding this underrepresented area of computing to the curriculum generates interest that may lead them to consider career opportunities in AI governance and policy. We see evidence of this phenomenon in our post-survey data, with 38\% of the respondents expressing interest in an AI policy career path compared to only 18\% before the module. As such, we see broad appeal for the continuation of the module in some form.

Nevertheless, the module falls short in truly connecting \textit{AI ethics and policy} to the technical content of the machine learning course. Questions such as ``\textit{how} do you prevent hallucination in large language models?'', or ``\textit{how} do you maintain data privacy in an AI system'' are left unaddressed by the two lectures. In part, this gap is understandable given that these questions represent still open research frontiers to which entire graduate seminars or academic careers are dedicated. Yet, as Skirpan et al. \cite{skirpan_ethics_2018} emphasize, integrating ethics into technical courses \textit{in situ} and not as a separate, independent entity is critical to underscore the importance of ethics to the field of computing. Given that machine learning and AI are not introductory computing subjects, the difficulty here arises in distilling complex, state-of-the-art algorithms to tackle ethical concerns like privacy or fairness into manageable programming examples or student projects. We see this as a still open research problem for the AI education community.

We do not, however, view this limitation as a repudiation of the purpose of the AI policy module. Instead, the module should be conceptualized as encompassing the core `must-know' topics, providing the context necessary to act as a springboard for further class activities or assignments. Moreover, if the module is incorporated towards the beginning of the term, the instructor could build in time for the discussion of the ethical challenges associated with specific algorithms introduced in lectures, or use ethical considerations to motivate the teaching of newer algorithms that aim to mitigate harms. Likewise, students could be asked, as part of an assignment or class project, to attempt to design and develop a technical solution to an ethical AI problem using the techniques and algorithms taught in the class. 

\subsection*{In-Class Game}

As evidenced by the student feedback, the in-class game \textit{Congress vs. Evil Inc.} proved to be a engaging centerpiece for the entire module. The juxtaposition of Congress against an `evil' technology company is purposefully hyperbolized, as is the cynical (and simplistic) use of `bribe cards' with which \textit{Evil Inc.} can bribe members of Congress to vote against key AI regulation. The aim of this exaggeration is to make the game appealing and easy to play while preserving the key tension between what regulations voters want and what regulation is (or isn't) actually enacted following the policymaking and lobbying process. Each of the three scenarios described in Appendix B were inspired by current events and propose  regulations that have been or could have been considered by legislators or regulators. These real-world connections were underscored during a post-game debriefing, encouraging students to take away the concept of competing policy interests, instead of the hyperbolized details of the game. 

Mechanically, however, the game could be further refined to keep students engaged throughout the gameplay and to balance the power of each of the game's three roles. In particular, the students noted that the motivations for how voters should vote were undefined, leading fewer members of Congress to be voted out of office as a result of voter dissatisfaction. In future iterations of the game, it may prove helpful to provide voters with player cards defining each voter's level of engagement with AI policy developments and their attitudes towards stricter AI regulation. In this case, each voter would have a predetermined motivation for how to vote. The distribution of voter preferences could be made to mirror public opinion polling data.

Gameplay balancing notwithstanding, we posit that the in-class game was a particular success because it had very little learning curve, emulating key voter-Congress-corporation dynamics while also being fun to play. The lack of out-of-class preparation required of students positions this style of game to be a more efficient active learning opportunity compared to a debate or role-play, which often require students to do independent research beforehand. In this way, our \textit{Congress vs. Evil Inc.} game is a effective proof-of-concept for integrating (unconventional) active learning in teaching computing ethics, even with more complex topics like AI policy.

\subsection*{Pedagogical Considerations}

Reflecting on our experience piloting this module in a graduate class, we conclude with a few suggestions regarding logistical or pedagogical choices. First, it should be noted that our university's lecture slots run for 1 hour and 15 minutes. As such, it is likely that the module would need to be split into three lectures for a 1-hour lecture schedule, in order to retain all key content. However, this may be preferable, as the in-class game could be separated from the ``Regulating Artificial Intelligence'' lecture and shifted to the third lecture day by itself, along with a culminating class debriefing or reflection.   

The other category of pedagogical considerations relates to how the module fits into the larger technical course from a participation and assessment perspective. Unfortunately, our module pilot saw between 25\% and 50\% less student attendance than normal, which is likely the result of students having already taken the course final exam and attendance not being tracked during the two module lectures. While the scheduling of the module right after the university's week-long Thanksgiving break is likely another contributing factor, instructors should consider how to require or incentivize attendance and participation in the class discussions through a roster check or a reflection assignment related to the module. Including an assignment as part of the module would also emphasize that this content is an integral part of the course and should not be treated as an afterthought.

\section*{Conclusion}

This paper details a two-lecture AI policy module for an introductory machine learning class, showing that content on AI ethics and AI policy can be effectively integrated into technical courses. While student ethical attitudes may not be easily changed—perhaps because they are solidified outside of the classroom—, students reported increased confidence in their ability to discuss issues of AI regulation and implement AI policies in their work after taking part in the module. Critically, the module increased student interest in following news developments about AI regulation and convinced some students to consider AI policy as a potential career path. The combination of lecture- and discussion-driven context with in-class activities and games to connect abstract concepts with real-world applications highlights a new path forward for designing AI ethics courses with an active learning pedological approach.

\newpage

\vspace{4\baselineskip}\vspace{-\parskip} 
\footnotesize 
\bibliographystyle{IEEEtranN} 
\bibliography{ASEEpaper}

\begin{thebibliography}{30}
\providecommand{\natexlab}[1]{#1}
\providecommand{\url}[1]{#1}
\csname url@samestyle\endcsname
\providecommand{\newblock}{\relax}
\providecommand{\bibinfo}[2]{#2}
\providecommand{\BIBentrySTDinterwordspacing}{\spaceskip=0pt\relax}
\providecommand{\BIBentryALTinterwordstretchfactor}{4}
\providecommand{\BIBentryALTinterwordspacing}{\spaceskip=\fontdimen2\font plus
\BIBentryALTinterwordstretchfactor\fontdimen3\font minus \fontdimen4\font\relax}
\providecommand{\BIBforeignlanguage}[2]{{%
\expandafter\ifx\csname l@#1\endcsname\relax
\typeout{** WARNING: IEEEtranN.bst: No hyphenation pattern has been}%
\typeout{** loaded for the language `#1'. Using the pattern for}%
\typeout{** the default language instead.}%
\else
\language=\csname l@#1\endcsname
\fi
#2}}
\providecommand{\BIBdecl}{\relax}
\BIBdecl

\bibitem[Jobin et~al.(2019)Jobin, Ienca, and Vayena]{jobin_global_2019}
A.~Jobin, M.~Ienca, and E.~Vayena, ``\BIBforeignlanguage{en}{The global landscape of {AI} ethics guidelines},'' \emph{\BIBforeignlanguage{en}{Nature Machine Intelligence}}, vol.~1, no.~9, pp. 389--399, Sep. 2019.

\bibitem[Parinandi et~al.(2024)Parinandi, Crosson, Peterson, and Nadarevic]{parinandi_investigating_2024}
S.~Parinandi, J.~Crosson, K.~Peterson, and S.~Nadarevic, ``Investigating the politics and content of {US} {State} artificial intelligence legislation,'' \emph{Business and Politics}, vol.~26, no.~2, pp. 240--262, Jun. 2024.

\bibitem[Aler~Tubella et~al.(2023)Aler~Tubella, Mora-Cantallops, and Nieves]{aler_tubella_how_2023}
A.~Aler~Tubella, M.~Mora-Cantallops, and J.~C. Nieves, ``How to teach responsible {AI} in {Higher} {Education}: challenges and opportunities,'' \emph{Ethics and Information Technology}, vol.~26, Dec. 2023.

\bibitem[exe()]{exec_order}
Exec. Order No. 14110, 88 CFR 75191. 2023. Available: \url{https://www.federalregister.gov/documents/2023/11/01/2023-24283/safe-secure-and-trustworthy-development-and-use-of-artificial-intelligence}.

\bibitem[Weichert et~al.(2025)Weichert, Kim, Zhu, and Eldardiry]{weichert_i_2025}
J.~Weichert, D.~Kim, Q.~Zhu, and H.~Eldardiry, ``'{Do} {I} {Have} to {Take} {This} {Class}?': {A} {Review} of {Ethics} {Requirements} in {Computer} {Science} {Curricula},'' in \emph{Proceedings of the {ACM} {Technical} {Symposium} on {Computer} {Science} {Education}}.\hskip 1em plus 0.5em minus 0.4em\relax Pittsburgh, PA, USA: Association for Computing Machinery, 2025.

\bibitem[wei({\natexlab{a}})]{weichert_eval}
J. Weichert, M. Cheng, D. Kim, Q. Zhu, and H. Eldardiry. ``Exploring the Evolution of AI Ethics in CS Ethics Courses,'' Forthcoming.

\bibitem[Bonwell and Eison(1991)]{bonwell_active_1991}
C.~C. Bonwell and J.~A. Eison, \emph{Active Learning: Creating Excitement in the Classroom}, ser. {ASHE}-{ERIC} Higher Education Research Report.\hskip 1em plus 0.5em minus 0.4em\relax School of Education and Human Development, George Washington Univ, 1991, no.~1.

\bibitem[Freeman et~al.(2014)Freeman, Eddy, McDonough, Smith, Okoroafor, Jordt, and Wenderoth]{freeman}
S.~Freeman, S.~L. Eddy, M.~McDonough, M.~K. Smith, N.~Okoroafor, H.~Jordt, and M.~P. Wenderoth, ``Active learning increases student performance in science, engineering, and mathematics,'' \emph{Proceedings of the National Academy of Sciences}, vol. 111, no.~23, pp. 8410--8415, 2014.

\bibitem[Vogt et~al.(2005)Vogt, Atwong, and Fuller]{vogt}
G.~Vogt, C.~Atwong, and J.~Fuller, ``Student assessment of learning gains (salgains): An online instrument,'' \emph{Business Communication Quarterly}, vol.~68, no.~1, pp. 36--43, 2005.

\bibitem[Shaw et~al.(2019)Shaw, Yang, Nash, Pigg, and Grim]{shaw_knowing_2019}
T.~J. Shaw, S.~Yang, T.~R. Nash, R.~M. Pigg, and J.~M. Grim, ``Knowing is half the battle: {Assessments} of both student perception and performance are necessary to successfully evaluate curricular transformation,'' \emph{PLoS ONE}, vol.~14, no.~1, 2019.

\bibitem[Brown et~al.(2024)Brown, Xie, Sarder, Fiesler, and Wiese]{brown_teaching_2024}
N.~Brown, B.~Xie, E.~Sarder, C.~Fiesler, and E.~S. Wiese, ``\BIBforeignlanguage{en}{Teaching {Ethics} in {Computing}: {A} {Systematic} {Literature} {Review} of {ACM} {Computer} {Science} {Education} {Publications}},'' \emph{\BIBforeignlanguage{en}{ACM Transactions on Computing Education}}, vol.~24, no.~1, pp. 1--36, Mar. 2024.

\bibitem[Fiesler et~al.(2020)Fiesler, Garrett, and Beard]{fiesler_what_2020}
C.~Fiesler, N.~Garrett, and N.~Beard, ``\BIBforeignlanguage{en}{What {Do} {We} {Teach} {When} {We} {Teach} {Tech} {Ethics}?: {A} {Syllabi} {Analysis}},'' in \emph{\BIBforeignlanguage{en}{Proceedings of the 51st {ACM} {Technical} {Symposium} on {Computer} {Science} {Education}}}.\hskip 1em plus 0.5em minus 0.4em\relax Portland OR USA: ACM, Feb. 2020, pp. 289--295.

\bibitem[Burton et~al.(2017)Burton, Goldsmith, Koenig, Kuipers, Mattei, and Walsh]{burton_ethical_2017}
E.~Burton, J.~Goldsmith, S.~Koenig, B.~Kuipers, N.~Mattei, and T.~Walsh, ``\BIBforeignlanguage{en}{Ethical {Considerations} in {Artificial} {Intelligence} {Courses}},'' \emph{\BIBforeignlanguage{en}{AI Magazine}}, vol.~38, no.~2, pp. 22--34, Jun. 2017.

\bibitem[Furey and Martin(2019)]{furey_ai_2019}
H.~Furey and F.~Martin, ``{AI} education matters: a modular approach to {AI} ethics education,'' \emph{AI Matters}, vol.~4, no.~4, pp. 13--15, Jan. 2019.

\bibitem[Garrett et~al.(2020)Garrett, Beard, and Fiesler]{garrett_more_2020}
N.~Garrett, N.~Beard, and C.~Fiesler, ``\BIBforeignlanguage{en}{More {Than} "{If} {Time} {Allows}": {The} {Role} of {Ethics} in {AI} {Education}},'' in \emph{\BIBforeignlanguage{en}{Proceedings of the {AAAI}/{ACM} {Conference} on {AI}, {Ethics}, and {Society}}}.\hskip 1em plus 0.5em minus 0.4em\relax New York NY USA: ACM, Feb. 2020, pp. 272--278.

\bibitem[Borenstein and Howard(2021)]{borenstein_emerging_2021}
J.~Borenstein and A.~Howard, ``\BIBforeignlanguage{en}{Emerging challenges in {AI} and the need for {AI} ethics education},'' \emph{\BIBforeignlanguage{en}{AI and Ethics}}, vol.~1, no.~1, pp. 61--65, Feb. 2021.

\bibitem[Raji et~al.(2021)Raji, Scheuerman, and Amironesei]{raji_you_2021}
I.~D. Raji, M.~K. Scheuerman, and R.~Amironesei, ``\BIBforeignlanguage{en}{You {Can}'t {Sit} {With} {Us}: {Exclusionary} {Pedagogy} in {AI} {Ethics} {Education}},'' in \emph{\BIBforeignlanguage{en}{Proceedings of the 2021 {ACM} {Conference} on {Fairness}, {Accountability}, and {Transparency}}}.\hskip 1em plus 0.5em minus 0.4em\relax Virtual Event Canada: ACM, Mar. 2021, pp. 515--525.

\bibitem[Larson et~al.(2016)Larson, Mattu, Kirchner, and Angwin]{larson_how_2016}
\BIBentryALTinterwordspacing
J.~Larson, S.~Mattu, L.~Kirchner, and J.~Angwin, ``How {We} {Analyzed} the {COMPAS} {Recidivism} {Algorithm},'' 2016. [Online]. Available: \url{https://www.propublica.org/article/how-we-analyzed-the-compas-recidivism-algorithm}
\BIBentrySTDinterwordspacing

\bibitem[Schiff et~al.(2020)Schiff, Biddle, Borenstein, and Laas]{schiff_whats_2020}
\BIBentryALTinterwordspacing
D.~Schiff, J.~Biddle, J.~Borenstein, and K.~Laas, ``\BIBforeignlanguage{en}{What's {Next} for {AI} {Ethics}, {Policy}, and {Governance}? {A} {Global} {Overview}},'' in \emph{\BIBforeignlanguage{en}{Proceedings of the {AAAI}/{ACM} {Conference} on {AI}, {Ethics}, and {Society}}}.\hskip 1em plus 0.5em minus 0.4em\relax New York NY USA: ACM, Feb. 2020, pp. 153--158. [Online]. Available: \url{https://dl.acm.org/doi/10.1145/3375627.3375804}
\BIBentrySTDinterwordspacing

\bibitem[wei({\natexlab{b}})]{weichert_policies}
J. Weichert, Q. Zhu, D. Kim, and H. Eldardiry. ``Perceptions of AI Ethics Policies Among Scientists and Engineers in Policy-Related Roles: An Exploratory Investigation,'' \emph{Digital Society} (2025). Forthcoming.

\bibitem[noa()]{noauthor_playable_nodate}
\BIBentryALTinterwordspacing
``Playable {City} {Sandbox}: {How} (not) to get hit by a self-driving car.'' [Online]. Available: \url{https://www.playablecity.com/projects/playable-city-sandbox-how-not-to/}
\BIBentrySTDinterwordspacing

\bibitem[Awad et~al.(2018)Awad, Dsouza, Kim, Schulz, Henrich, Shariff, Bonnefon, and Rahwan]{awad_moral_2018}
E.~Awad, S.~Dsouza, R.~Kim, J.~Schulz, J.~Henrich, A.~Shariff, J.-F. Bonnefon, and I.~Rahwan, ``\BIBforeignlanguage{en}{The {Moral} {Machine} experiment},'' \emph{\BIBforeignlanguage{en}{Nature}}, vol. 563, no. 7729, pp. 59--64, Nov. 2018.

\bibitem[Haq et~al.(2024)Haq, Zhu, Hui, and Tyson]{haq_deepfakes}
E.-U. Haq, Y.~Zhu, P.~Hui, and G.~Tyson, ``History in making: Political campaigns in the era of artificial intelligence-generated content,'' in \emph{Companion Proceedings of the ACM Web Conference 2024}, ser. WWW '24.\hskip 1em plus 0.5em minus 0.4em\relax New York, NY, USA: Association for Computing Machinery, 2024, p. 1115–1118.

\bibitem[Floridi and Cowls(2019)]{floridi_unified_2019}
L.~Floridi and J.~Cowls, ``\BIBforeignlanguage{en}{A {Unified} {Framework} of {Five} {Principles} for {AI} in {Society}},'' \emph{\BIBforeignlanguage{en}{Harvard Data Science Review}}, Jun. 2019.

\bibitem[Kim et~al.(2023)Kim, Zhu, and Eldardiry]{kim_exploring_2023}
D.~Kim, Q.~Zhu, and H.~Eldardiry, ``Exploring {Approaches} to {Artificial} {Intelligence} {Governance}: {From} {Ethics} to {Policy},'' in \emph{2023 {IEEE} {International} {Symposium} on {Ethics} in {Engineering}, {Science}, and {Technology} ({ETHICS})}.\hskip 1em plus 0.5em minus 0.4em\relax West Lafayette, IN, USA: IEEE, May 2023, pp. 1--5.

\bibitem[Auld et~al.(2022)Auld, Casovan, Clarke, and Faveri]{auld_governing_2022}
G.~Auld, A.~Casovan, A.~Clarke, and B.~Faveri, ``\BIBforeignlanguage{en}{Governing {AI} through ethical standards: learning from the experiences of other private governance initiatives},'' \emph{\BIBforeignlanguage{en}{Journal of European Public Policy}}, vol.~29, no.~11, pp. 1822--1844, Nov. 2022.

\bibitem[Floridi(2019)]{floridi_translating_2019}
L.~Floridi, ``\BIBforeignlanguage{en}{Translating {Principles} into {Practices} of {Digital} {Ethics}: {Five} {Risks} of {Being} {Unethical}},'' \emph{\BIBforeignlanguage{en}{Philosophy \& Technology}}, vol.~32, no.~2, pp. 185--193, Jun. 2019.

\bibitem[eu_()]{eu_ai_act}
EU AI Act, Regulation (EU) 2024/1689. 2024. Available: \url{https://eur-lex.europa.eu/eli/reg/2024/1689/oj/eng}.

\bibitem[Weichert and Eldardiry(2024)]{weichert_computer_2024}
J.~Weichert and H.~Eldardiry, ``Computer {Science} {Student} {Attitudes} {Towards} {AI} {Ethics} and {Policy}: {A} {Preliminary} {Investigation},'' in \emph{Proceedings of the 2024 {IEEE} {International} {Symposium} on {Technology} and {Society}}, Puebla, Mexico, 2024.

\bibitem[Skirpan et~al.(2018)Skirpan, Beard, Bhaduri, Fiesler, and Yeh]{skirpan_ethics_2018}
M.~Skirpan, N.~Beard, S.~Bhaduri, C.~Fiesler, and T.~Yeh, ``\BIBforeignlanguage{en}{Ethics {Education} in {Context}: {A} {Case} {Study} of {Novel} {Ethics} {Activities} for the {CS} {Classroom}},'' in \emph{\BIBforeignlanguage{en}{Proceedings of the 49th {ACM} {Technical} {Symposium} on {Computer} {Science} {Education}}}.\hskip 1em plus 0.5em minus 0.4em\relax Baltimore Maryland USA: ACM, Feb. 2018, pp. 940--945.

\end{thebibliography}


\newpage

\section*{Appendix A}

\begin{table}[H]
    \centering
    \begin{tabular}{l|p{13cm}}
         \multicolumn{1}{c|}{\textbf{Question Type}} & \multicolumn{1}{c}{\textbf{Question}} \\
         \midrule
         \multicolumn{2}{c}{\textit{Section 1: AI Ethics}} \\
         \midrule
         Free Response & In one sentence, how would you define the term \textit{ethics}? \\
         Likert & In general, I think existing AI tools are ethical. \\
         Likert & I believe that most developers of AI tools design their AI systems with ethics in mind. \\
         Likert & I worry about the ethical impact of *current* AI technology. \\
         Likert & I worry about the ethical impact of *future* AI technology. \\
         Likert & In general, I think I act ethically when I use or create AI tools. \\
         Free Response & Are there any particular ethical concerns or impacts of AI technology that you are concerned about? \\
         \midrule
         \multicolumn{2}{c}{\textit{Section 2: AI Policy}} \\
         \midrule
         Free Response & In one sentence, how would you define \textit{AI policy}? \\
         Likert & I believe AI technologies are currently adequately regulated by the government. \\
         Likert & The government and private companies should do more to protect *users* from potential harms of AI technology. \\
         Likert & The government and private companies should do more to protect *society* from potential harms of AI technology. \\
         Likert & I can have a robust discussion with friends or peers about AI regulation. \\
         Likert & I plan to follow news about government regulation of technology and/or AI in the future. \\
         Likert & I feel confident in my ability to apply ethical principles to my work related to AI. \\
         Likert & I feel confident in my ability to implement policies regarding AI in my work. \\
         Likert & My future job will probably require me to be generally knowledgable about AI policy. \\
         Likert & I am interested in AI policy and regulation as a potential career path. \\
         \midrule
         \multicolumn{2}{c}{\textit{Section 3: AI Policy Module Feedback} (\textit{Post-Survey} only)} \\
         \midrule
         Free Response & Overall, how would you rate these two lectures (``AI Ethics'' and ``AI Policy'')? \\
         Likert & The content of the lectures was interesting. \\
         Likert & The lectures were engaging and interactive. \\
         Likert & I liked the level of interactivity in these lectures compared to previous lectures. \\
         Likert & The lectures were a good use of class time. \\
         Likert & The in-class simulation (mafia game) was enjoyable. \\
         Likert & The in-class simulation (mafia game) was helpful for me to connect with the lecture content. \\
    \end{tabular}
    \caption{Pre- and Post-Survey Questions}
    \label{tab:survey-questions}
\end{table}

\newpage
\section*{Appendix B}

\begin{table}[H]
    \centering
    \begin{tabular}{l|>{\raggedright}p{2.5cm}|>{\raggedright}p{2cm}|p{8.5cm}}

         \textbf{Scenario} & \textbf{Topic} & \textbf{Key Issue} & \textbf{Proposed Regulations} \\
         \midrule
         Scenario 1 & AI-targeted advertising & Protection of underage users & \begin{itemize}
             \item Social media ban for children under 18
             \item Users can request a copy of their own data
             \item Users must be notified about when and how AI is used
             \item Ban on using AI to target advertising
         \end{itemize}\\ 
        \midrule
         Scenario 2 & Autonomous vehicles & Testing and acceptable harm & \begin{itemize}
             \item Require companies to monitor their autonomous fleet from a central control center
             \item Require autonomous vehicles to have a manual override feature
             \item Prohibit autonomous vehicles from being publicly released until they pass a rigorous final inspection
             \item Ban autonomous vehicles in all cases
         \end{itemize}\\
         \midrule
         Scenario 3 & LLMs and `general-purpose' AI & Disclosure of model information & \begin{itemize}
             \item Prohibit LLMs from training on user data (prompts)
             \item Require LLMs to disclose their general model architecture and training process
             \item Require LLMs to notify law enforcement if the user refers to illegal activity
             \item Prohibit LLMs from accessing the live internet (i.e. require a training cutoff)
         \end{itemize}\\
    \end{tabular}
    \caption{\textit{Congress vs. Evil Inc.} Game Scenarios}
    \label{tab:game-scenarios}
\end{table}

\newpage
\section*{Appendix C}

\begin{figure}[H]
    \setlength{\belowcaptionskip}{1\baselineskip}
    \centering
    \begin{subfigure}{.6\textwidth}
        \includegraphics[width=1\linewidth]{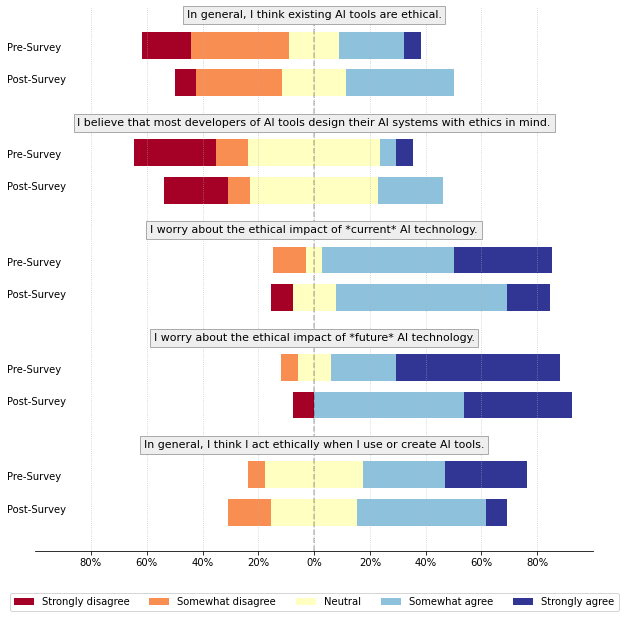}
        \caption{Attitudes towards \textit{AI ethics} statements.}
        \label{fig:ethics-likert}
    \end{subfigure}
    \begin{subfigure}{.6\textwidth}
        \includegraphics[width=1\linewidth]{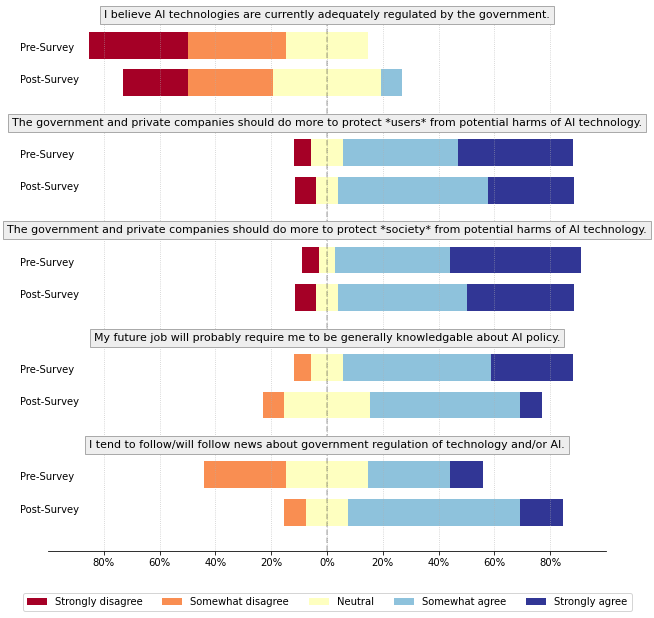}
        \caption{Attitudes towards \textit{AI policy} statements.}
        \label{fig:policy-likert}
    \end{subfigure}
    \caption{Student attitudes towards AI ethics and AI policy statements before and after the AI policy module.}
\end{figure}

\end{document}